# Designing Socially Intelligent Virtual Companions

Han Yu, Zhiqi Shen, Qiong Wu, and Chunyan Miao
{han.yu, zqshen, wuqi0005, ascymiao}@ntu.edu.sg
Nanyang Technological University

**Abstract.** Virtual companions that interact with users in a socially complex environment require a wide range of social skills. Displaying curiosity is simultaneously a factor to improve a companion's believability and to unobtrusively affect the user's activities over time. Curiosity represents a drive to know new things. It is a major driving force for engaging learners in active learning. Existing research work pays little attention in curiosity. In this paper, we enrich the social skills of a virtual companion by infusing curiosity into its mental model. A curious companion residing in a Virtual Learning Environment (VLE) to stimulate user's curiosity is proposed. The curious companion model is developed based on multidisciplinary considerations. The effectiveness of the curious companion is demonstrated by a preliminary field study.

**Keywords:** Socially intelligent companion, human-computer interaction, curiosity.

## 1. Introduction

In order to establish a close and long-term relationship with their users, virtual companions have attracted major research interest from the field of artificial intelligence. A number of companion-related research projects are going on, including Companion [1], SERA [2], CLASSiC [3], LIREC [4], COGNIRON [5], SEMAINE [6], etc. It has been agreed by researchers that the artificial companion should be accessible to users over a long period of time in a socially complex environment. Hence, all these projects emphasize the importance of social skills in virtual companions.

To imbue social intelligence into virtual companions, various facets of social abilities, such as conversation [7], emotion [8], memory [9], and trust [10], have been considered by researchers. However, these facets overlooked the aspect of social motivation which is essential for promoting a life filled with novelty. For example, an emotionally intelligent conversational companion without curiosity may talk the users out of a bad mood, but cannot proactively seek out potentially interesting news or activities to engage the users in an active lifestyle.

Curiosity is the force driving active exploration in human beings [15]. A healthy dose of curiosity has been found to result in the development of capabilities and more importantly, creativity [12], [13]. Graesser et al. found that a curious peer can stimulate another's curiosity by pointing out novel information to him/her through social interactions [14]. Hence, curiosity can be a very important social skill of virtual companions to promote the curiosity of users over long-term interactions and potentially enhance a life with novelty.

In this paper, we enrich the social skills of virtual companions by constructing a curiosity model for them. The proposed virtual curious companion resides in a Virtual Learning Environment (VLE), which is an ideal platform to study its effectiveness in stimulating the curiosity in users.

The remaining parts of the paper are organized as follows: Section 2 gives an overview of psychological studies on human curiosity and the state-of-the-art of curious agent design. In Section 3, our proposed socially intelligent Curious Companion (CC) is presented, with a usage scenario to demonstrate the reasoning process of the CC and its interactions with the user. The results from a field study to measure the effectiveness of the proposed CC in terms of interest retention are discussed in Section 4. Section 5 concludes the paper and lists important future research issues.

## 2. Related Work

In psychological studies, the concept of curiosity in human being can be divided into two dimensions [13]: 1) diversive curiosity, which is aroused when people are bored or hungry for information; and 2) specific curiosity, which is aroused when new information are surprising or conflicting with one's existing understanding. From these definitions, curiosity is partially determined by a person's innate characteristics and the external stimuli he/she receives from the environment. The innate urge to be curious about one's surroundings is primarily driven by his/her personality - more specifically, the propensity to be curious [19]. This characteristic is found in psychological studies to determine the intensity of diversive curiosity and one's attention to novelty which, in turn, drives the process of novelty discovery in the information one receives. The novelty that has been discovered in this process will likely be the external trigger for specific curiosity in the subject matter and may cause further in-depth exploration into this specific domain. The resulting enhanced understanding gained from this exercise will make subsequent encounters with the same concepts appear less novel to the person. The relationships between various factors affecting human curiosity are summarized in the cognitive map shown in Figure 1.

Designing curious agents is a research problem that has attracted attentions from many researchers. However, the primary aim of previous research work on CCs is about making curiosity an intrinsic drive for the agents, rather than their users, to explore. For instance, Schmidhuber [16] has demonstrated the effectiveness of curiosity in directing the agents to explore dynamic environments. Reinforcement learning and intrinsic rewards were used in that study to direct the curious agent to refine its model of the environment. Marsland et al. [17] incorporated curiosity into robots to equip them with novelty seeking behaviours which help them with exploration. Macedo and Cardosa [18] infused the concept of surprise into CCs to induce further exploration into the surprising areas. Saunders [19] uses CCs to study the use of computational curiosity modelling to help software agents discover novelty in creative works (e.g. image patterns).

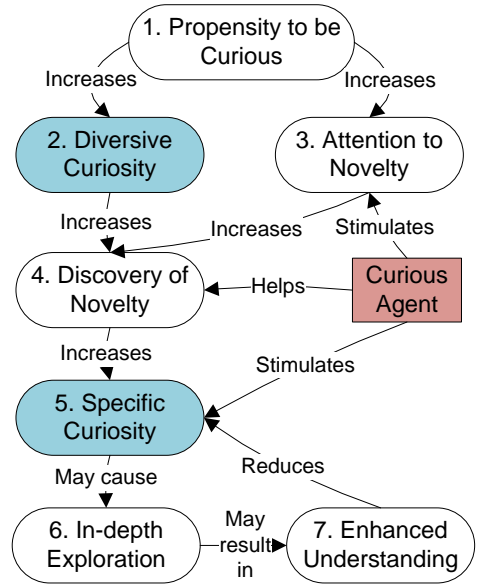

**Figure 1. The influence of the human propensity to be curious on the two dimensions of curiosity, their effect in driving exploration [19], and how the proposed CC helps in this process.**

While these works all confirm the important relationship between curiosity, motivation, learning and creativity, they do not aim at developing these qualities in human users to enhance their learning experience and long term cognitive development. Merrick [20] has made preliminary investigations into this direction by designing curious robotic toys whose movement patterns change when reconfigured to study its effect on stimulating players' curiosity. However, the interactions between curious robot and the users are limited and no study results on its effectiveness have been published at the time of writing of this paper. In the next section, we discuss important considerations when designing a curious learning companion to help users be curious when studying science subjects.

## 3. A Simple Curious Companion Model

Depending on one's knowledge and past experience, what appears to be novel or surprising to one learner might be a familiar fact for another. Therefore, a CC must tailor its stimulation conditions to the learning progress of different learners even if the underlying concepts being taught are the same. In addition, whenever curiosity stimulation is decided to be necessary, the level of stimulation that a learner can tolerate must be taken into consideration. If the stimulus issued by the CC is too complex, too novel or too irritating, anxiety or revulsion might be aroused from the learner instead of the desired curiosity.

As an open-ended environment, a VLE provides ample time for exploration by the learners once their curiosity is aroused. In such an environment, exchanging questions with a large number of peers is much easier for a learner than in a classroom. As text chatting is the prevailing medium of message exchange in VLEs, discussion is made even easier for people who are shy to speak up in front of other (which is quite a common phenomenon in oriental cultures). The novel virtual objects and the sense of immersion (sensory, actional and symbolic [21]) provide a readily available intrinsic reward for exploration and discovery to the learners. These opportunities offered by the VLEs make it an ideal platform for studying the use of CCs to stimulate learners' curiosity and develop their creativity over the long term. With these considerations in mind, we propose the following architectural design for the curious companions.

**3.1 The Proposed Curious Companion Architecture**

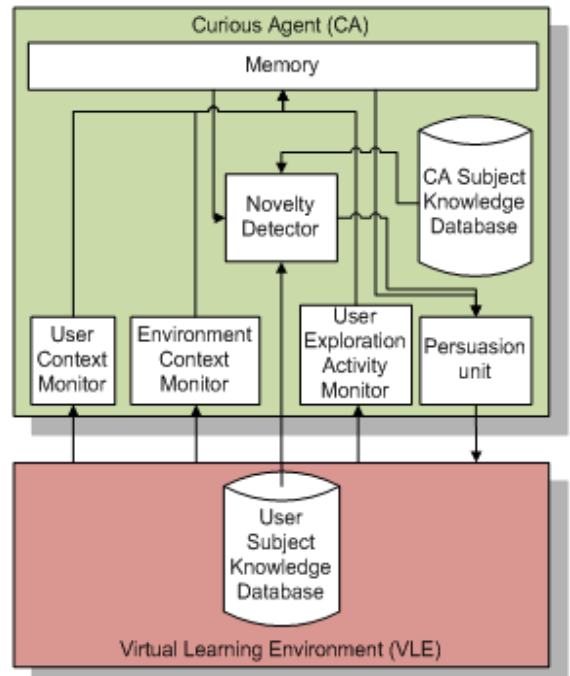

**Figure 2. The design architecture of the Proposed CC [29].**

The aforementioned considerations are summarized into a simple curious companion architecture as shown in Figure 2. Its functional modules can be divided into three generic categories: 1) a perception module, which is responsible for sensing the necessary domain of interest and collecting relevant data to support the subsequent decisions made by the CC; 2) a cognition module, which contains the main algorithms for achieving the design objectives of the CC; and 3) a curiosity stimulation module, which is responsible for interacting with the learner.

**User Context**: The CC should be able to monitor the learner's context for the purpose of personalizing novelty detections and tailoring unobtrusive curiosity stimulations. In a VLE, a user's context may include information from 3 different aspects:

1) *The learner's current topic of learning and his/her progress*: as each individual learner may have different progress in learning, it is unsuitable to sweepingly mark certain knowledge concepts as novel to the learner and raise it to the learner's attention as a teacher does in a classroom. To facilitate manipulation by both human beings and the CC, the knowledge representation mechanism must satisfy the following requirements: 1) it can represent knowledge in a symbolic way to facilitate human understanding; 2) it supports the computational analysis of a problem; and 3) it can describe complex relationships among many concepts which are useful for learning. In view of these considerations, we have chosen fuzzy cognitive maps (FCMs) as a medium for knowledge representation in the proposed CC.

   Fuzzy Cognitive Map (FCM) is a fuzzy-graph structure which can simulate the complex systems in the world through causes, effects, and the causal relationships in between [22]. A FCM can be defined in the form of *(C, R, W)*, where $C = \{C_1, C_2, \ldots, C_N\}$ is the set of concepts. These concepts are represented as nodes in a FCM graph. The set of causal relationships among these concepts is denoted by $R = \{R_{11}, R_{12}, \ldots, R_{NN}\}$, where $R_{ij}$ represents the strength and nature of influence on $C_j$ by $C_i$. The causal relations can be compactly represented by a connection matrix $\mathbf{W} = [w_{ij}], 1 \leq i \leq N, 1 \leq j \leq N$. Thus, a learner b's current subject knowledge, $K_b$, is represented as:

$$K_b = \langle \begin{array}{c} \{C_i^{(b)}\}_{i=1}^N, \\ \mathbf{W}^{(b)} = [w_{ij}], 1 \leq i \leq N, 1 \leq j \leq N \end{array} \rangle \quad (1)$$

2) *The learner's propensity to be curious and tolerance to stimuli*: people differ in their intrinsic level of diversive curiosity that drives each of them to initiate their own exploration. This is an important parameter for the CC to decide whether to issue curiosity stimuli or not and the intensity and complexity of the stimulation that are required to change the learner's attitude about exploration. As it is a relatively stable personality trait, a person's propensity to be curious can be assessed using questionnaires such as the curiosity and exploration inventory (CEI) proposed by Kashdan et al. [15]. A learner's curiosity profile can be denoted as follows:

$$cp^{(b)} = \{r_i^{(b)}\}_{i=1}^M \quad (2)$$

where $cp^{(b)}$ is the curiosity profile of a learner *b*, $r_i^{(b)}$ is learner *b*'s response to self-assessment question number *i*. The curiosity profile to which a user is classified dictates what curiosity stimulation strategies should be adopted by the CC throughout the subsequent interactions with the user.

3) *The learner's possible current mood*: from time to time, we all experience periods of moodiness during which nothing seems to be worth exploring for us. Recent studies in behavioural science confirm that there is a positive correlation between positive emotions and curiosity [23]. Thus, knowing a user's likely mood is useful for making curiosity stimulations unobtrusive. However, as a person's mood changes during learning activities, it is not practical to resolve to lengthy questionnaires for assessment. Therefore, non-intrusive measuring techniques such as using mouse clicks and keyboard press data [24] offer promising solutions in this case. The historical mouse and keyboard actions of a learner *b* is denoted as a collection of records as:

$$\text{mk}^{(b)} = \{\text{ac}_t^{(b)}\} \quad (3)$$

where $\text{ac}_t^{(b)}$ is the type of action (e.g. left mouse click, right mouse click, mouse movement, key press, etc.) performed by learner *b* at time *t*.

**Environment Context**: Learning concepts in a VLE tend to be embedded in virtual objects, virtual characters or learning activities in the form of:

$$A = \{p, \{C_i\}_{i=1}^N\} \quad (4)$$

where A is the learning activity, *p* denotes the properties of *A* (e.g. name, objectives, background information, etc.) and $\{C_i\}$ is the set of subject knowledge that is covered in *A*. Thus, it is important for the CC to be able to extract this piece of information from the environment to know when a learner is coming into contact with certain new knowledge. In addition, in order to make sense of a user's current context, his/her location in the virtual environment, which is intuitively represented by (x, y, z) in a three dimensional virtual world, needs to be constantly tracked. Actions that the user can perform in a VLE to participate in various learning activities should be clearly defined to enable the CC to monitor if the user is in need of prompt to continue exploring.

**Novelty Detection**: During the learning-by-exploration process in a VLE, it is important for the CC to be able to discern which piece of knowledge is novel to a particular learner in his/her current context. This requires the CC to possess a novelty detector which makes use of the information collected by the user context monitor and the environment monitor to decide whether the knowledge a user is about to encounter in a certain location in a VLE is conflicting with his/her current understanding of the topic. In addition, a form of representation of a learner's knowledge that is both easy for human beings to construct and understand, and also simple enough for computers to store and make inference is vital for the novelty detector to work.

In this study, we adopt the classic view from psychological research that regards curiosity as a motivational state which elicits interest from the learner by presenting them with "something new" or "something surprising" [25]. In the proposed CC model, we consider the "something new" to be the concepts that a learner has not yet encountered in the VLE and the "something surprising" to be knowledge embedded in the virtual learning activity that contradicts what the learner has learnt (or misunderstood) from past experience in the VLE. To facilitate the detection of novelty, not only the acquired concepts should be recorded in the learner's knowledge base, basic relationships among these concepts should also be contained in the knowledge representation to support the level of learning required by the target group of our study – which is high school students.

With FCMs as the knowledge model for both the learner and the proposed CC, novelty detection can be accomplished by the CC according to the following procedure (assuming the same set of symbols is used for both the learners and the CC to denote individual concepts in the subject matter):

1) The CC computers the intersection of $C_L$ and $C_{CC}$ as $C_I = C_L \cap C_{CC}$, where $C_L$ is the set of concepts in the learner's current understanding FCM and is the $C_{CC}$ is the set of concepts in the CC's knowledge base (which is the preloaded FCM representation of the subject knowledge designed by a subject expert), assuming the same set of symbols are used by both the learner and the CC to represent to corresponding concepts. If $C_I \neq C_{CC}$, $C_{new} = C_{CC} - C_L$, where $C_{new}$ is a set containing the concepts that the learner has not yet encountered in the VLE.
2) Remove all the rows and columns that correspond with concepts in $C_{new}$ from the matrix $W_{CC}$, which denotes the causal relationship among the concepts in the CC's knowledge base, to form $W'_{CC}$. In this way, the FCM representing the CC's subject knowledge is reduced to the same dimension as the FCM representing the learner's current knowledge.
3) Fuzzify the magnitude of the entries in $W_{CC}$ and $W'_{CC}$. Compare $W'_{CC}$ with WL (which denotes the causal relationship among the concepts in the learner's knowledge base). If $W'_{CC}[i,j] \neq W_L[i,j]$, add [i, j] into the set $R_S$ which denotes the set of surprises to the learner.
4) Determine the current location of the learner's avatar in the VLE, and the set of learning activities $A = \{A_1, A_2, ..., A_H\}$ in the vicinity. For each learning activity $A_a \in A$, acquire the set of subject knowledge concepts, CC, embedded in it. If $\exists c_k \in C_A \land (c_k \in C_{new} \lor k \in R_S)$, then mark $A_a$ as potentially novel for the learner.

**Decision to Stimulate Curiosity**: A learner may become curious about the newly encountered learning activity on his/her own and initiate exploration. In this situation, external stimulus by the CC would be redundant and even be considered annoying. Thus, the CC should monitor the learner's activity to decide whether stimulating his/her curiosity is necessary in order to make the stimuli unobtrusive.

After detecting the potentially novel learning activities near to the learner's avatar, the prototype CC waits for duration τ for the learner's action. This is an intuitive attempt to give the learner enough time to see if he/she could become curious about one of the novel learning activities and start to take part in it. For the current CC prototype, the value for τ is determined heuristically analysing the learner's past mouse movement pattern in the context of the location of the learn-er's avatar. While a learner can use mouse movements to control the direction his/her avatar is facing, and therefore, view the environment from different angles, only mouse clicks and keyboard strokes can move the avatar or activate events in the test-bed VLE. Thus, in our attempt to quantify the value of $\tau$ for each learner, we consider the average time lapse between keyboard strokes or mouse clicks $T$. Thus, $\tau = nT, n \in Z^+$. After a time period of $\tau$, if the learner still has not engaged in, or is moving his/her avatar away from, the potentially novel learning activities, the CC will randomly select one of the novel learning activities and generate a prompt (currently in the form of a question) to suggest that the learner investigate it.

The rationale for using this rather static heuristic approach is due to the lack of studies conducted in the area of virtual worlds to investigate the correlation between people's mouse and keyboard usage pattern and their likely state of the mind. Ideally, if a mapping function $F$ exists:

$$F: mk^{(b)} \rightarrow P_m \langle \alpha, \nu, \delta \rangle \qquad (5)$$

where $P_m$ is a learner's likely mood profile corresponding to his/her historical mouse and keyboard usage pattern, $\langle \alpha, \nu, \delta \rangle$ denote his/her level of arousal, valence and dominance respectively, which are popular dimensions used to measure a person's mood in psychological studies [24]. Once more findings from on the influence of mood on a person's level of curiosity is made available, a wide range of agent learning technologies can be applied to make their stimulation decisions more personalized and appropriate.

**Curiosity Stimulation Strategy**: When the CC decides to stimulate a learner's curiosity, it needs to be equipped with persuasive strategies to make the stimulation effective. The most common means of stimulation is the use of thought-provoking questions that highlight surprising facts to the learner. The composition of such sentences by the CC needs to rely on the information from the novelty detection unit (i.e. what is the new or surprising concept to a particular user), employing appropriate persuasion strategies [28] and medium of delivery. Apart from this, highlighting complex problems or uncertainty may also to be effective ways of stimulating one's curiosity [26]. Moreover, the way the stimuli are constructed must also take into account the learner's tolerance level for them so as not to overwhelm the learner and cause anxiety (e.g. the stimulus should make exploring certain new concepts look appealing and enjoyable rather than making the process look overly difficult for someone who does not enjoy challenges). The mapping process from a learner's curiosity profile to a possible stimulation strategy can be represented by f:

$$f: cp^{(b)} \rightarrow ss_k \langle c, m, Ps \rangle \qquad (6)$$

where $ss$ is a stimulation strategy that contains specifications for the complexity of the stimulus ($c$), the learner's preferred medium of stimulation ($m$) and the appropriate persuasion strategy ($Ps$) for him/her. Tailored suggestions are one of the persuasion strategies for the CC. The possible suggestions in the current CC prototype for each learning activity have been pre-constructed and incorporated into the companion. However, it is possible to summarize the way questions are construct-ed to elicit curiosity from a user into a set of templates (e.g. "Would you like to learn something about" + something_new + "? You can explore it by participating in" + learning_activity + "here." or "There may be other ways of explaining the concept" + something_surprising + ", would you like to see how it is done in" + learning_activity + "?", etc.). This way, it is possible to dynamically generate the suggestions. Should additional stimulation strategies become available, more templates could be designed to accommodate them.

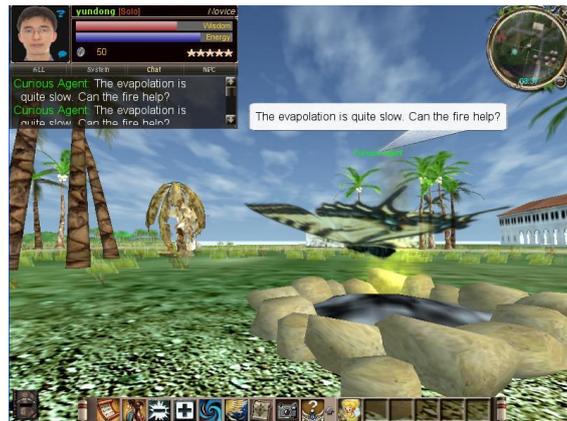

**Figure 3. The CC in our test-bed VLE is embodied by a butterfly that follows each individual learner.**

The current prototype CC is implemented using the Goal Net Methodology [30], [31] and embodied by a little butterfly, as shown in Figure 3, which follows each individual user while he/she explores the test-bed VLE.

### 3.2 A Usage Scenario for the Prototype CC

Through exploring the CoS VLE, a learner *L* has acquired some knowledge on the topic of the transport systems in plants. After each learning activity ends, he updates the FCM representation of his current understanding of the subject by selecting from a predefined set of concept nodes and specifying the relationships among these concepts using the interface provided by a teachable agent from our previous study [27] in the VLE as shown in Figure 4.

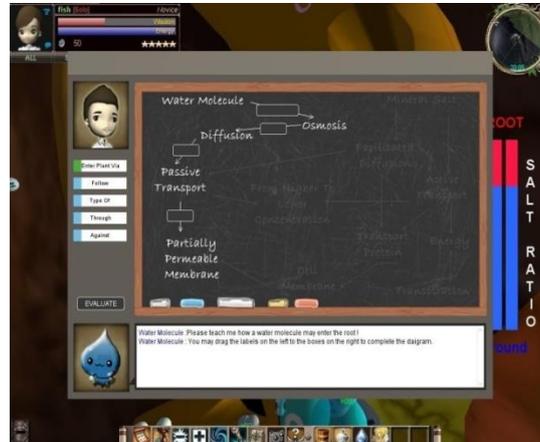

**Figure 4. Interface in the VLE for the learner to construct a FCM representation of his/her own current understanding of the subject incrementally.**

At the current stage, his knowledge on the subject matter is illustrated by the FCM shown in Figure 5. Figure 6 illustrates the FCM representation of the CC's subject knowledge. The parts in Figure 5 and 6 that are highlighted in red denote the differences between these two FCMs. As can be seen, the learner has only encountered concepts C1 to C8 and misunderstood the relationship between C4 and C7 (he thinks higher salt concentration in the soil will facilitate the plant to absorb water through osmosis).

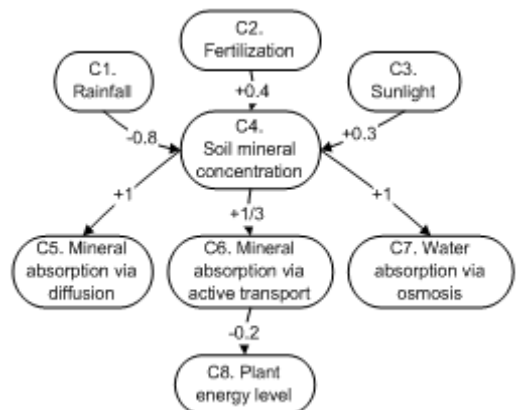

**Figure 5. An example of a learner's knowledge represented by a FCM.**

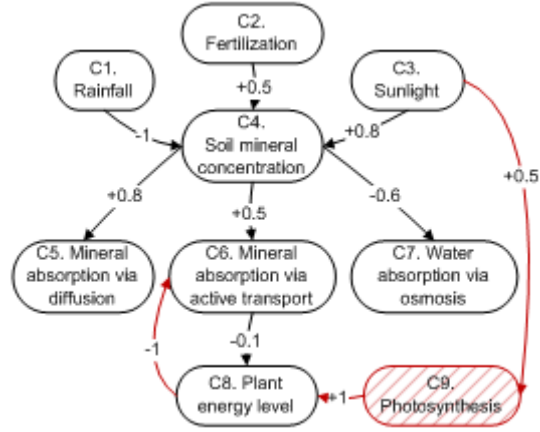

**Figure 6. An example of the CC's subject knowledge represented by a FCM.**

By going through Step 1 of novelty detection method in the previous section, the CC identifies the set of new concepts for the learner based on his current learning progress and stores them into a set $C_{new}$:

$$C_{new} = \{C_9\} \qquad (6)$$

The weight matrix for the learner's knowledge FCM and the CC's knowledge FCM are denoted as $W_L$ and $W_{CC}$ respectively.

$$W_L = \begin{bmatrix} 0 & 0 & 0 & -0.8 & 0 & 0 & 0 & 0 \\ 0 & 0 & 0 & +0.4 & 0 & 0 & 0 & 0 \\ 0 & 0 & 0 & +0.3 & 0 & 0 & 0 & 0 \\ 0 & 0 & 0 & 0 & +1 & +1/3 & +1 & 0 \\ 0 & 0 & 0 & 0 & 0 & 0 & 0 & 0 \\ 0 & 0 & 0 & 0 & 0 & 0 & 0 & -0.2 \\ 0 & 0 & 0 & 0 & 0 & 0 & 0 & 0 \\ 0 & 0 & 0 & 0 & 0 & 0 & 0 & 0 \end{bmatrix}$$

$$W_{CC} = \begin{bmatrix} 0 & 0 & 0 & -1 & 0 & 0 & 0 & 0 & 0 \\ 0 & 0 & 0 & +0.5 & 0 & 0 & 0 & 0 & 0 \\ 0 & 0 & 0 & +0.8 & 0 & 0 & 0 & 0 & +0.5 \\ 0 & 0 & 0 & 0 & +0.8 & +0.5 & -0.6 & 0 & 0 \\ 0 & 0 & 0 & 0 & 0 & 0 & 0 & 0 & 0 \\ 0 & 0 & 0 & 0 & 0 & 0 & 0 & -0.1 & 0 \\ 0 & 0 & 0 & 0 & 0 & 0 & 0 & 0 & 0 \\ 0 & 0 & 0 & 0 & 0 & -1 & 0 & 0 & 0 \\ 0 & 0 & 0 & 0 & 0 & 0 & 0 & +1 & 0 \end{bmatrix}$$

Following Step 2 in the proposed novelty detection method, $W_{CC}$ can be reduced to $W'_{CC}$:

$$W'_{CC} = \begin{bmatrix} 0 & 0 & 0 & -1 & 0 & 0 & 0 & 0 \\ 0 & 0 & 0 & +0.5 & 0 & 0 & 0 & 0 \\ 0 & 0 & 0 & +0.8 & 0 & 0 & 0 & 0 \\ 0 & 0 & 0 & 0 & +0.8 & +0.5 & -0.6 & 0 \\ 0 & 0 & 0 & 0 & 0 & 0 & 0 & 0 \\ 0 & 0 & 0 & 0 & 0 & 0 & 0 & -0.1 \\ 0 & 0 & 0 & 0 & 0 & 0 & 0 & 0 \\ 0 & 0 & 0 & 0 & 0 & -1 & 0 & 0 \end{bmatrix}$$

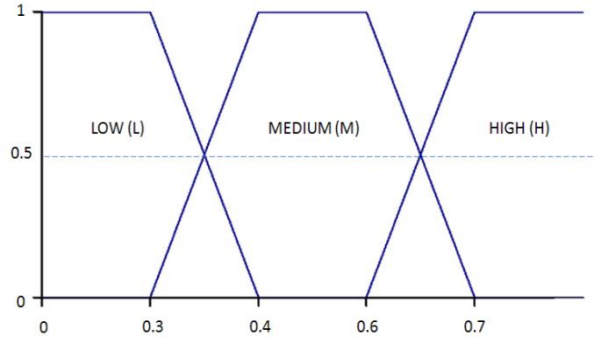

**Figure 7. The prototype CC trying to stimulate a learner's curiosity to explore a learning activity.**

Comparing the exact decimal values in the two weight matrices would not produce meaningful results. Therefore, fuzzification of these values is needed to make sense of the really substantial differences that may appear surprising to a learner. In this prototype, the weight values are in the range of 0 to 1. This range is divided into three fuzzy sets: low (L), medium (M) and high (H) as illustrated in Figure 7. The membership functions corresponding to these three fuzzy sets are:

$$\beta(x) = \begin{cases} 1, & 0 < x < 0.3 \\ -10x + 4, & 0.3 \leq x < 0.4 \end{cases}$$
$$\theta(x) = \begin{cases} 10x - 3, & 0.3 \leq x < 0.4 \\ 1, & 0.4 \leq x < 0.6 \\ -10x + 7, & 0.6 \leq x < 0.7 \end{cases} \quad (7)$$
$$\gamma(x) = \begin{cases} 10x - 6, & 0.6 \leq x < 0.7 \\ 1, & x \geq 0.7 \end{cases}.$$

To simplify the subsequent comparisons, the fuzzy set with the highest membership function evaluation for a given $w_{ij}$ value is used to denote the relationship $r_{ij}$ between concepts $i$ and $j$:

$$r_{ij} = \mathrm{argmax}_{x \in [0,1]} \langle \beta(x)L, \theta(x)M, \gamma(x)H \rangle \quad (8)$$

Following this step, the WL and $W'_{CC}$ are further refined into $W_{L(f)}$ and $W'_{CC(f)}$:

$$W_{L(f)} = \begin{bmatrix} 0 & 0 & 0 & -H & 0 & 0 & 0 & 0 \\ 0 & 0 & 0 & +M & 0 & 0 & 0 & 0 \\ 0 & 0 & 0 & +L & 0 & 0 & 0 & 0 \\ 0 & 0 & 0 & 0 & +H & (+0.5L, +0.5M) & +H & 0 \\ 0 & 0 & 0 & 0 & 0 & 0 & 0 & 0 \\ 0 & 0 & 0 & 0 & 0 & 0 & 0 & -L \\ 0 & 0 & 0 & 0 & 0 & 0 & 0 & 0 \\ 0 & 0 & 0 & 0 & 0 & 0 & 0 & 0 \end{bmatrix}$$

$$W'_{CC(f)} = \begin{bmatrix} 0 & 0 & 0 & -H & 0 & 0 & 0 & 0 \\ 0 & 0 & 0 & +M & 0 & 0 & 0 & 0 \\ 0 & 0 & 0 & +H & 0 & 0 & 0 & 0 \\ 0 & 0 & 0 & 0 & +H & +M & -M & 0 \\ 0 & 0 & 0 & 0 & 0 & 0 & 0 & 0 \\ 0 & 0 & 0 & 0 & 0 & 0 & 0 & -L \\ 0 & 0 & 0 & 0 & 0 & 0 & 0 & 0 \\ 0 & 0 & 0 & 0 & 0 & -H & 0 & 0 \end{bmatrix}$$

$$W_{result} = \begin{bmatrix} 0 & 0 & 0 & 0 & 0 & 0 & 0 & 0 \\ 0 & 0 & 0 & 0 & 0 & 0 & 0 & 0 \\ 0 & 0 & 0 & 1 & 0 & 0 & 0 & 0 \\ 0 & 0 & 0 & 0 & 0 & 1 & 1 & 0 \\ 0 & 0 & 0 & 0 & 0 & 0 & 0 & 0 \\ 0 & 0 & 0 & 0 & 0 & 0 & 0 & 0 \\ 0 & 0 & 0 & 0 & 0 & 0 & 0 & 0 \\ 0 & 0 & 0 & 0 & 0 & 1 & 0 & 0 \end{bmatrix}$$

Comparing the differences in $W_{L(f)}$ and $W'_{CC(f)}$, the element $W_{result}$ [3, 4], $W_{result}$ [4, 6], $W_{result}$ [4, 7] and $W_{result}$ [8, 6] are set to the value 1, which means that the causal relationship between concepts $C_4$ and $C_7$ as well as that between $C_8$ and $C_6$ in these two FCMs are different.

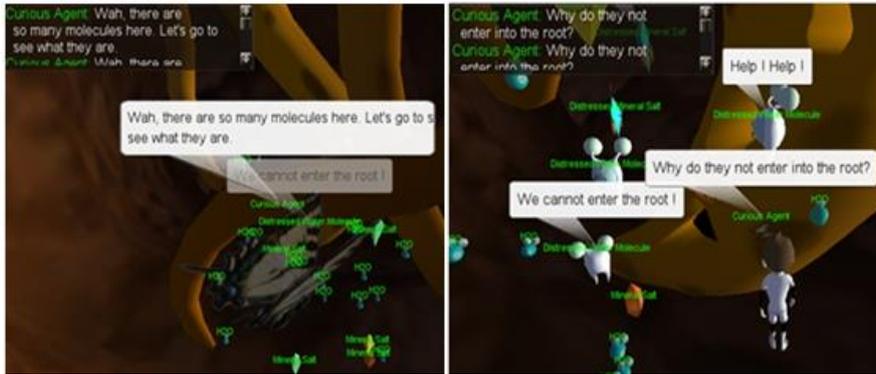

**Figure 8. The prototype CC trying to stimulate a learner's curiosity to explore a learning activity.**

Therefore, the 2-tuples ⟨3,4⟩, ⟨4,6⟩, ⟨4,7⟩ and ⟨8,6⟩ are inserted in the set $R_S$ which contains all the surprises (conflicts in understanding) for the learner based on his current understanding of the subject matter. $R_S = \{\langle 3,4\rangle, \langle 4,6\rangle, \langle 4,7\rangle, \langle 8,6\rangle\}$. At that moment, location of the learner's avatar in the CoS VLE is near the root of a plant where learning activities include helping salt molecules enter the root structure which teaches the learner the diffusion process ($A_1\{C_5\}$); and helping the water molecules enter the root structure which teaches the learner about osmosis ($A_2\{C_7\}$). Following step 3 of the proposed novelty detection method, the concept of diffusion ($C_5$) in $A_1$ does not belong to either $C_{new}$ or $R_S$, while the concept of osmosis is denoted as $C_7$ in the learner's knowledge FCM which does not belong to $C_{new}$ but, nevertheless, belongs to $R_S$.

Therefore, after determining that the learner has no intention to explore the current location, the CA issues a textual prompts in the form of questions to the learner as shown in the Figure 8 to stimulate the learner's curiosity about participating in the learning activity $A_2\{C_7\}$ at this virtual venue.

## 4. Field Study

The CoS VLE has been studied in the Catholic High School in Singapore to study its effectiveness. Although the CC was not singled out as a focus of the study, its effectiveness in terms of retaining students' interest in the VLE has been assessed as part of the study.

### 4.1 Method of the Study

Two groups of Secondary Two level (equivalent to Grade 8 in the North American high school system) students who are rated as having average academic abilities by their teachers, were selected to participate in the study. The topic of their study was transport systems in plants, which was chosen from their science class curriculum. Before the study commences, both groups have not learnt the chosen topic at the secondary school level (but they did encounter this topic during their primary school years). They were given a pre-test on the subject matter to gauge their understanding. During the study, which consists of four separate sessions, one group learnt this topic by using the CoS VLE, and the control group went through normal classroom teaching. At the end of the study, both groups sat for a post-test on the topic and the group that used the CoS VLE also gave their feedback on the system in the form of a survey.

### 4.2 Analysis of Results

At the current stage, the post-test scripts are still being marked and tallied by the school teachers. Therefore, we will use the survey results for analysis of the performance of the proposed CC in this application environment.

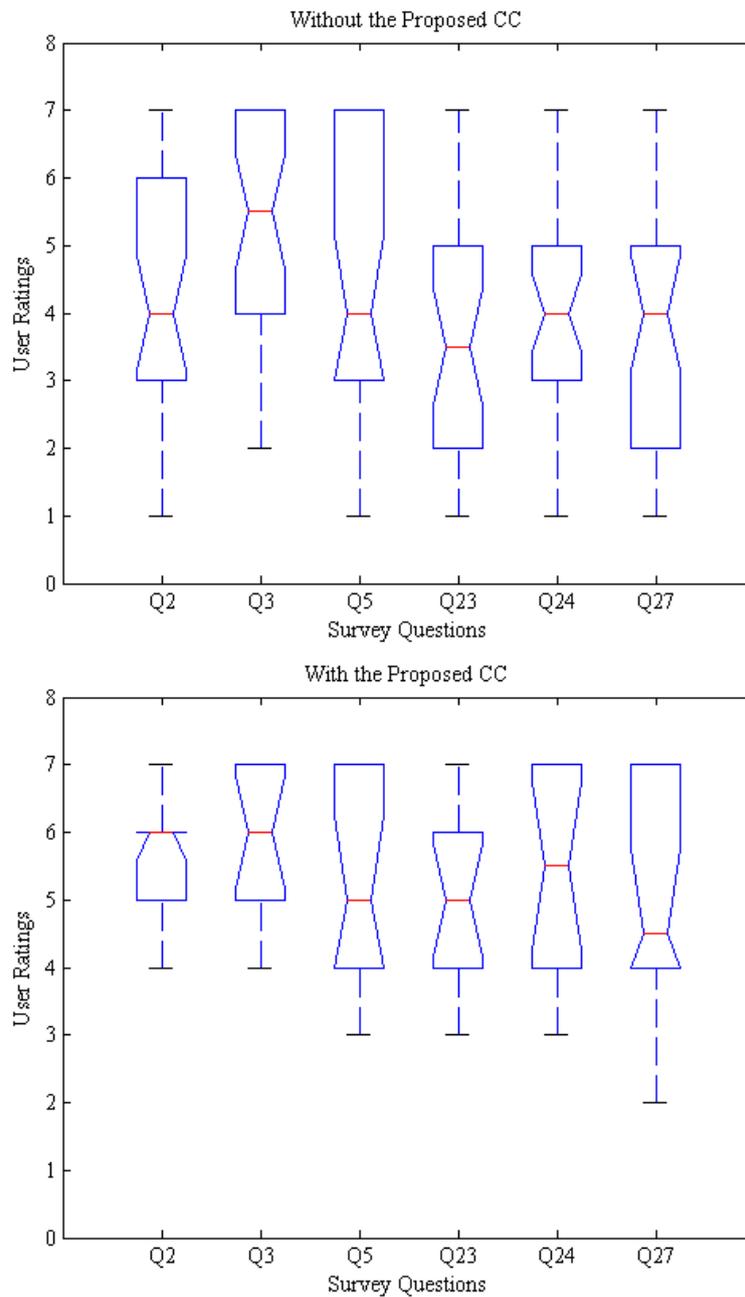

**Figure 9. The distributions of ratings for survey questions concerning the ease of use of FCMs as the student knowledge representation tool and the effectiveness of the proposed CC in terms of interest retention by the two groups of users.**

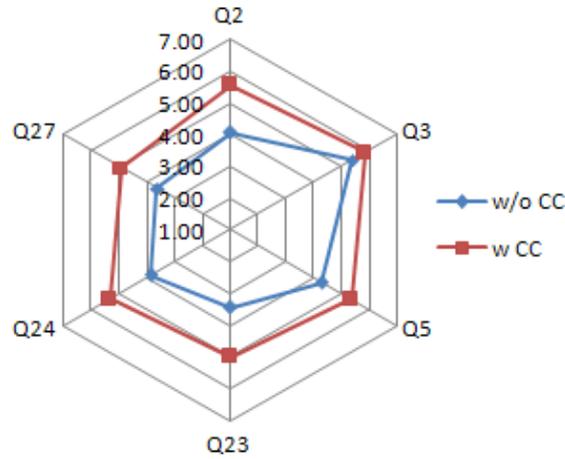

**Figure 10. The average ratings for survey questions concerning the ease of use of FCMs as the student knowledge representation tool and the effectiveness of the proposed CC in terms of interest retention by the two groups of users.**

The survey consists of 28 rating questions assessing various aspects of the system. The ratings are in the range of 1 – 7 with 1 being completely disagree and 7 being completely agree. Three of these questions (Q2, Q3 and Q5) measure the students' perception of how effective their interest in exploring learning activities in the VLE has been retained throughout the sessions. Another three of these questions (Q23, Q24, Q27) measure the students' perceived ease of use for the user knowledge representation format in the VLEs. In the version of VLE without the proposed CC, the students input their knowledge by adding "if…then…" rules to a rule base. In the version with the proposed CC, the students input their knowledge by modifying a FCM.

As shown in Figure 9 and Figure 10, comparing both the distribution as well as the average of ratings between the group using the CoS VLE without the proposed the CC and the group using the same VLE with the proposed CC, students from the latter group perceived that their interest in the exploring the learning activities has been better retained. An average rating of 5.6 out of 7 over Q2, Q3 and Q5 has been achieved by the group using the proposed CC while the group that did not use the proposed CC only achieved an average rating of 4.45 out of 7 over the same questions. This represents a 25.8% improvement. Students also prefer to use FCM as a knowledge representation mechanism in the VLE than simple rule bases. This validates our design choice of using FCM for the students to record the knowledge they have acquired in the VLE, which also simplifies the process of novelty detection in the proposed CC.

A two-sample unpooled *t*-test with unequal variances is performed on the following hypothesis:

*Null Hypothesis* ($H_0$): there is no statistically significant difference between the average rating scores from the group with the proposed CC and the group without it.

*Alternative Hypothesis* ($H_1$): there is statistically significant difference between the average rating scores from the group with the proposed CC and the group without it.

**Table 1.** The 2-tailed, two-sample unpooled Student *t*-test with unequal variances for the average rating scores from the two groups of students on the effectiveness of the two versions of the same VLE in terms of interest retention.

| Group | Statistics | Survey Questions on Interest Retention |
|---|---|---|
| Group without the proposed CC | Sample Size | 33 |
| | Mean Score | 4.45 |
| | Variance | 1.351 |
| Group with the proposed CC | Sample Size | 30 |
| | Mean Score | 5.60 |
| | Variance | 1.753 |
| | *t*-value | 2.897 *(Significant)* |
| At 95% confidence level, $t_{critical}$ = [-1.997, 1.997] | | |

As shown in Table 1, the test statistic $t = 2.897 \notin [-1.997, 1.997]$. Thus we reject $H_0$ with 95% confidence. This shows that there is statistically significant evidence based on this survey that the proposed CC does contribute to retaining the students' interest in exploring the learning activities in a VLE.

Comments from the students show that they are generally surprised by the novelty of the proposed CC. The proposed CC, in itself, might qualify as something new and surprising to them. Some of them have been observed to be curious about the CC and interested in exploring the range of capabilities of the CC itself by bringing it to a wide range of different locations in the VLE.

## 5. Conclusions and future work

In this paper, we proposed a socially intelligent curious virtual companion to stimulate user's curiosity through social interactions. We have discussed important design considerations that should be taken in order to make the CC practical. Based on the model, a prototype CC has been implemented in our test-bed VLE. Field studies on its effectiveness in retaining learners' interest in exploring the learning activities in a VLE has yield promising results.

Nevertheless, at the current stage, many further refinements to the current CC model are needed in order to achieve the long term goal of the proposed CC to foster a life of interest for the users in pursuit of life-long learning.

1) *Computational curiosity models*: being curious is distinctly a human experience that is not simply binary. A human level of curiosity has varying intensities and provides an intrinsic drive to his/her actions. Being able to quantify the level of curiosity a person may be experiencing is an important first step towards influencing his/her attitudes. We believe curiosity brings a new dimension in research on various intelligent agents/robots such as learning, negotiation, persuasion, inference, trust agents and various virtual companions.

2) *Curiosity stimulation design patterns*: in complex tasks such as learning in a VLE, the stimulations a CC provides to a user could potentially be in multiple media (e.g. text, audio, animation sequences, etc.) to appeal to people with different preferences. How to effectively stimulate a user's curiosity the key problem that will ultimately determine whether a CC is successful or not. Although techniques for stimulating curiosity in a classroom setting has been proposed from the educational research field, we believe this process holds more similarities to persuasion in a virtual environment. Effective curiosity inducing interaction patterns for a virtual world based environment catering to users with different personalities will be explored to enhance the usefulness of future CCs.

3) *Assessing the effectiveness of a curious companion*: the current practice of assessing the interest retention and motivation effectiveness of the CC based on user feedback can be subjective and inadequate. It is inherently difficult for a human being to quantify these concepts. Thus, we plan to look into the possibility of analysing the changes in the users' behaviour patterns under the influence of CCs to more objectively assess the effectiveness of a particular CC design. This will open up the opportunity to incorporate the design based research methodology [23] into the CC design process.

These issues will be addressed in our subsequent research work.

## Acknowledgements

This research is supported, in part, by the National Research Foundation, Prime Minister's Office, Singapore under its IDM Futures Funding Initiative and administered by the Interactive and Digital Media Programme Office.

## References

[1] E.M. Altmann and W.D. Gray, "Managing attention by preparing to forget," In Human Factors and Ergonomics Society Annual Meeting Proceedings, Cognitive Ergonomics, vol. 1(4), pp.152–155, 2000.

[2] A.D. Baddeley, "Is working memory still working?", European Psychologist, vol. 7(2), pp.85–97, 2002.

[3] V. Rieser and O. Lemon, "Learning and Evaluation of Dialogue Strategies for new Applications: Empirical Methods for Optimization from Small Data Sets," Computational Linguistics, vol. 37(1), pp.153–196, 2011.

[4] J.R. Anderson, "A Spreading Activation Theory of Memory," Journal of Verbal Learning and Verbal Behavior, vol. 22, 1983.

[5] J.W. Alba and L. Hasher, "Is memory schematic?" Psychological Bulletin, vol. 93, pp.203–231, 1983.

[6] R. Atkinson and R. Shiffrin, "Human memory: A Proposed System and its Control Processes," The Psychology of Learning and Motivation: Advances in Research and Theory, vol. 2, 1968.

[7] G. Salvi, F. Tesser, E. Zovato, and P. Cosi, "Cluster Analysis of Differential Spectral Envelopes on Emotional Speech," Eleventh Annual Conference of the International Speech Communication Association, 2010.

[8] H. Prendinger and M. Ishizuka, "The Empathic Companion: A Character-based interface that Addresses Users's Affective States," Applied Artificial Intelligence, vol. 19(3-4), pp.267-285, 2005.

[9] M. Lim, "Memory Models for Intelligent Social Companions," Human-Computer Interaction: The Agency Perspective, pp.241-262, 2012.

[10] M.U. Keysermann, R. Aylett, S. Enz, H. Cramer, C. Zoll, and P.A. Vargas, "Can I trust you? Sharing information with artificial companions," 11th International Conference on Autonomous Agents and Multiagent Systems (AAMAS), in press, 2012.

[11] K. Borowske, "Curiosity and Motivation-to-Learn," In Proceedings of Association of College and Research Libraries (ACRL) 12th National Conference, pp. 346-350, 2005.

[12] I. Li, A. Dey, and J. Forlizzi, "A Stage-Based Model of Personal Informatics Systems," ACM SIG CHI, 2010.

[13] C. Leuba, "A New Look and Curiosity and Creativity," The Journal of Higher Education, vol. 29(3), pp. 132-140, 1958.

[14] A. Graesser, N. Person, and J. Maglina, "Collaborative Dialogue Patterns in Naturalistic one-to-one Tutoring," Applied Cognitive Psychology, vol. 9(6), pp.495–522, 1995.

[15] T. Kashdan, P. Rose, and F. Fincham, "Curiosity and Exploration: Facilitating Positive Subjective Experiences and Personal Growth Opportunities," Journal of Personality Assessment, vol. 82(3), pp.291-305, 2004.

[16] J. Schmidhuber, "Curious Model-Building Control Systems," In Proceedings of the International Joint Conference on Neural Networks, vol. 2, pp.1458-1463, 1991.

[17] S. Marsland, U. Nehmzow, and J. Shapiro, "Novelty Detection for Robot Neotaxis," In International Symposium on Neural Computation, pp.554-559, 2000.

[18] L. Macedo and A. Cardosa, "Creativity and Surprise," In AISB'01 Symposium on AI and Creativity in Arts and Science, 2001.

[19] R. Saunders, "Supporting Creativity Using Curious Agents," In Workshop on Computational Creativity Support in the 27th Annual SIGCHI Conference on Human Factors in Computing Systems, 2009.

[20] K. Merrick, "Designing Toys That Come Alive: Curious Robots for Creative Play," In Proceedings of the 7th International Conference on Entertainment Computing (ICEC'08), pp.149-154, 2008.

[21] C. Dede, "Immersive Interfaces for Engagement and Learning," Science, vol. 323(5910), pp.66-69, 2009.

[22] B. Kosko, "Fuzzy Cognitive Maps," International Journal of Man-Machine Studies, vol. 24(1), pp.65-75, 1986.

[23] T.B. Kashdan and W.E. Breen, "Social Anxiety and Positive Emotions: A Prospective Examination of a Self-Regulatory Model with Tendencies to Suppress or Express Emotions as a Moderating Variable," Behavior Therapy 39(1), pp.1-12, 2008.

[24] P. Zimmermann, S. Guttormsen, B. Danuser, and P. Gomez, "Affective computing - a rationale for measuring mood with mouse and keyboard," International Journal on Occupational Safety and Ergonomics, vol. 9(4), pp.539-551, 2003.

[25] T.B. Kashdan and P.J. Silvia, "Curiosity and Interest: The Benefit of Thriving on Novelty and Challenge," Chapter 34 of Oxford Handbook of Positive Psychology, pp.367-375, 2009.

[26] W.H. Maw and E.W. Maw, "Nature of Curiosity in High-and Low-Curiosity Boys," Developmental Psychology, vol. 2(3), pp. 325-329, 1970.

[27] H. Yu, Y. Cai, Z. Shen, X. Tao and C. Miao, "Agents as Intelligent User Interfaces for the Net Generation," in Proceedings of the 15th International Conference on Intelligent User Interfaces (IUI'10), pp. 429-430, 2010.

[28] O-K. Harri and H. Marja, "Persuasive Systems Design: Key Issues, Process Model, and System Features," Communications of the Association for Information Systems, vol. 2, 2009.

[29] H. Yu, Z. Shen, C. Miao and A. Tan, "A Simple Curious Agent to help people be Curious," In Proceedings of the 10th International Conference on Autonomous Agent and Multiagent Systems (AAMAS'11), pp.1159-1160, 2011.

[30] H. Yu, Z. Shen, and C. Miao, "Intelligent Software Agent Design Tool Using Goal Net Methodology," In Proceedings of the 2007 IEEE/WIC/ACM International Conference on Intelligent Agent Technology (IAT'07), pp.43-46, 2007.

[31] H. Yu, Z. Shen, and C. Miao, "A Goal-Oriented Development Tool to Automate the Incorporation of Intelligent Agents into Interactive Digital Media Applications," ACM Computer in Entertainment (CIE) Magazine, vol. 6(2), 2008.